# Superconductor $Pb_{10-x}Cu_x(PO_4)_6O$ showing levitation at room temperature and atmospheric pressure and mechanism

Sukbae Lee,[1,a)] Jihoon Kim,[1] Hyun-Tak Kim,[2,3,b)] Sungyeon Im,[1] SooMin An,[1] and Keun Ho Auh[1,4]

[1]Quantum Energy Research Centre, Inc., Seoul 05822, South Korea
[2]ICT Basic Research Lab. ETRI, Daejeon 34129, South Korea
[3]Department of Physics, College of William & Mary, Williamsburg, VA 23185, USA
[4]Hanyang University, Seoul 04763, South Korea

a)Author to whom correspondence should be addressed : stsaram@qcentre.co.kr
b)Author to whom correspondence should be addressed : hkim22@wm.edu, hkim0711@snu.ac.kr

**Abstract**

A material called LK-99®, a modified-lead apatite crystal structure with the composition $(Pb_{10-x}Cu_x(PO_4)_6O$ (0.9<x<1.1)), has been synthesized using the solid-state method. The material exhibits the Ohmic metal characteristic of Pb(6s1) above its superconducting critical temperature, $T_c$, and the levitation phenomenon as Meissner effect of a superconductor at room temperature and atmospheric pressure below $T_c$. A LK-99® sample shows $T_c$ above 126.85°C (400K). We analyze that the possibility of room-temperature superconductivity in this material is attributed to two factors: the first being the volume contraction resulting from an insulator-metal transition achieved by substituting Pb with Cu, and the second being on-site repulsive Coulomb interaction enhanced by the structural deformation in the one-dimensional(D) chain ($Cu^{2+}(3d^9$ with one-hole carrier)$−O_{1/2}−Cu^{2+}(3d^9)$) along the c-axis) structure owing to superconducting condensation at $T_c$. The mechanism of the room-temperature $T_c$ is discussed by 1-D BR-BCS theory.

## I. INTRODUCTION

For over 110 years since Onnes's 1911 discovery of superconductivity, scientists have searched for room-temperature superconductors, materials that exhibit zero resistance to electrical current. Superconductors have been discovered in various metal elements and compound crystals. In 1986, a cuprate superconductor with a superconducting critical temperature, $T_c$, over 40 K was found.[1,2] In 2015, a hydride of $H_2S$ was shown to demonstrate $T_c \approx 203$ K at a pressure of 155 GPa.[3] In 2023, a $T_c$ of 294 K at 10 kbar was measured on a nitrogen-doped lutetium hydride;[4] moreover, a superconductor [LK-99®] of over $T_c$=300K was successfully synthesized (language: Korean).[5]

BCS(Bardeen-Cooper-Schrieffer) theory, which provides a microscopic explanation of superconductivity, was introduced in 1957.[6] The Brinkman-Rice(BR)-BCS theory accounting for $T_c$ of over room temperature was discovered in 2021.[7] Moreover, in 2021, a theory was published that predicts $T_c$ as approximately 10% of Fermi temperature.[8] In addition, whether the pairing symmetry of the superconducting energy gap in cuprate superconductors is *d*-wave or *s*-wave has been debated for over 30 years; recent studies suggest that it may lean towards *s*-wave symmetry.[9,10,11]

To discover a room-temperature superconductor, observing the emergence of a metallic phase through an insulator-to-metal transition (IMT) at temperatures higher than room temperature is crucial.[12,13] Moreover, to explain this phenomenon, the discovery of a new material that exhibits room-temperature superconductivity at atmospheric pressure, along with a comprehensive mechanism, is required.

In this paper, we present a synthesis method for a Cu-doped lead apatite (LA) superconductor with a Tc exceeding room temperature. We conduct levitation experiments and analyze the zero resistance properties of the material. Moreover, we unveil the mechanism by which the metal phase is formed through an insulator-to-metal transition (IMT) without undergoing a structural phase transition in LA. Furthermore, we provide a phase diagram for the superconductor. Finally, we briefly discuss the mechanism of room-temperature superconductivity based on the BR-BCS theory[7].

## II. RESULTS and DISCUSSIONS
### A. LK-99® : Synthesis
For sample manufacture, we synthesized $Pb_{10-x}Cu_x(PO_4)_6O$ (0.9<x<1.1), known as LK-99®, using the solid-state method. The raw materials used in the synthesis were PbO (Junsei, GR), $PbSO_4$ (Kanto, GR), Cu (Daejung, EP), and P (Junsei, EP).[5] The solid-state method is explained in Fig. 1. The process of sample synthesis comprises three steps. **Step 1:** To obtain Lanarkite $Pb_2(SO_4)O$ = PbO + $Pb(SO_4)$, PbO and $Pb(SO_4)$ powders were uniformly mixed in a ceramic crucible with a rate of 50% each. The mixed powder was heated in a furnace at 725°C for 24 hours in the presence of air [Fig. 1(b)]. During the heating process, the mixed materials underwent a chemical reaction, yielding lanarkite. **Step 2:** To synthesize $Cu_3P$, Cu and P powders were mixed in a crucible as per each component rate. The mixed powder was sealed in a crystal tube of 20 cm per gram with a vacuum of $10^{-3}$ torr [Fig. 1(a)]. The sealed tube containing the mixed materials was heated in a furnace at 550°C for 48 hours [Fig. 1(c)]. During this process, the mixed materials underwent a transformation and formed $Cu_3P$ crystals. **Step 3:** The Lanarkite and the $Cu_3P$ crystals were ground to make powder and mixed in a crucible. Then, the mixed powders were sealed in a crystal tube of a vacuum of $10^{-3}$ torr [Fig. 1(a)]. The sealed tube containing the mixed powder was heated in a furnace at 925°C for 5-20 hours [Fig. 1(d)]. During this process, the mixed powder reacted and transformed into the final material of $Pb_{10-x}Cu_x(PO_4)_6O$. The sulfur element present in $PbSO_4$ was evaporated during the reaction. The various shapes observed during the process are depicted in photos [Figs. 1(e-i)].

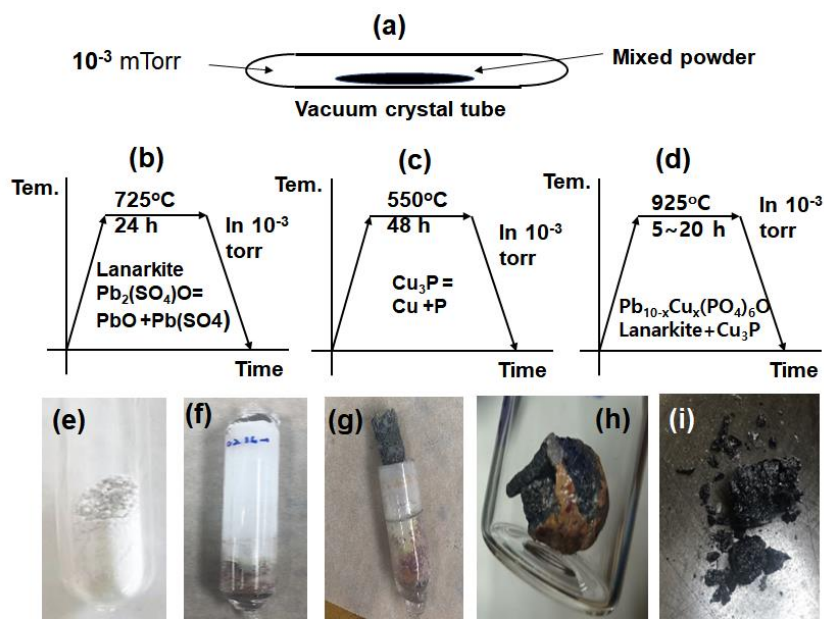

**FIG.1.** (a) Layout of sealed vacuum crystal tube with mixed power. (b), (c), (d) Heat treatment conditions of Lanarkite, $Cu_3P$, $Pb_{10-x}Cu_x(PO_4)O$ (0.9<x<1.1), respectively. (e) All ingredients premixed powder before reaction, appearing white to light gray. (f) Picture of the sealed sample after the reaction, (g) Sample removal procedure from the furnace, (h) A shape of the sample of the sealed quartz tube, (i) A sample shape in each process.

**B. LK-99® : Analysis of crystal structure**

The manufactured powder's crystallinity and structure were analyzed by X-ray diffraction (XRD) measurements and data fitting. Fig. 2(a) shows a comparison of XRD data measured with LK-99® sample 1 (over-doped material) and COD (Crystallography Open Database) data of the pure lead-apatite supported by QualX software. The XRD analysis of the sample showed multiple black peaks [Fig. 2(a)], indicating that it is a polycrystalline material. The XRD pattern of sample 1 closely matches that of modified-lead apatite (MLA) with slight peak shifts. However, the peak indicated by symbol A in sample 1 is shifted to a larger angle, and a new peak, indicated by symbol B, appears, suggesting a change in the lattice structure of sample 1, [Fig. 2(b)]. The shifts indicate a decrease of lattice constant, which is interpreted as evidence of volume contraction. Comparing the XRD pattern of sample 1 with MLA in the VESTA program confirms that sample 1 exhibits an MLA structure, [Fig. 2(c)]. Specifically, an LA structure is formed as $A_{10}(BO_4)C$, one of the frameworks of the hexagonal structure of element A.[14,15]

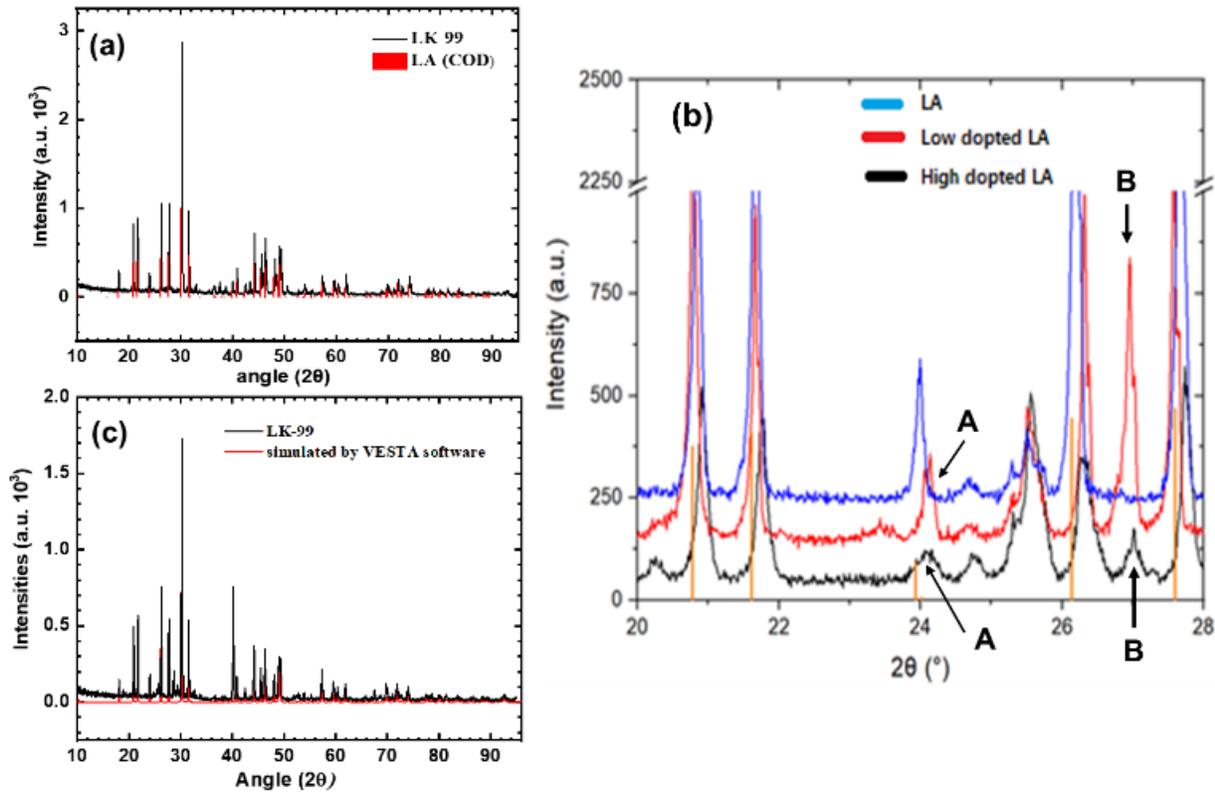

**FIG. 2.** (a) Comparison of XRD pattern measured with sample 1 and lead apatite data in Crystallography Open Database. (b) Magnified pattern shows a peak shift and a new peak. (c) The XRD pattern is compared with the modified apatite pattern obtained from BESTA program. It is closely fitted.

The XRD analysis reveals that sample 1 has a hexagonal structure of (P63/m, 176) with cell lattice parameters of a=9.843 Å and c=7.428 Å [Figs. 3(a) and 3(b)], while the LA's parameters are a=9.865 Å and c=7.431 Å. The volume of sample 1 shrinks by 0.48%, resulting from substituting Pb(M1) with Cu(M2). The phenomenon of shrinkage has already been studied in previous research on apatite materials.[16] Although it is classified as an LA structure, $Pb_{10}(PO_4)_6O$, is an insulator; conversely, Cu-doped LA, $Pb_{10-x}Cu_x(PO_4)_6O$, is a superconductor at room temperature and a metal above $T_c$. Moreover, it is structurally condensed by substituting $M1^{2+}$ (corresponding to $Pb1^{2+}$, at Pb(1) position in LA atomic classification in crystallography information file, cif.) at the black-colored position in Figs. 3(a) and 3(b) by $Cu^{2+}$ ion ($M2^{2+}$, one reddish-brown-colored among four Pb(1) positions in Fig. 3(a)). Furthermore, owing to a structural distortion at one-dimensional line of $Cu^{2+}-O_{1/2}-Cu^{2+}$ along the c-axis at $T_c$, as mentioned in Fig. 3(c), it becomes a superconductor.[7] Further details will be explained in a following section.

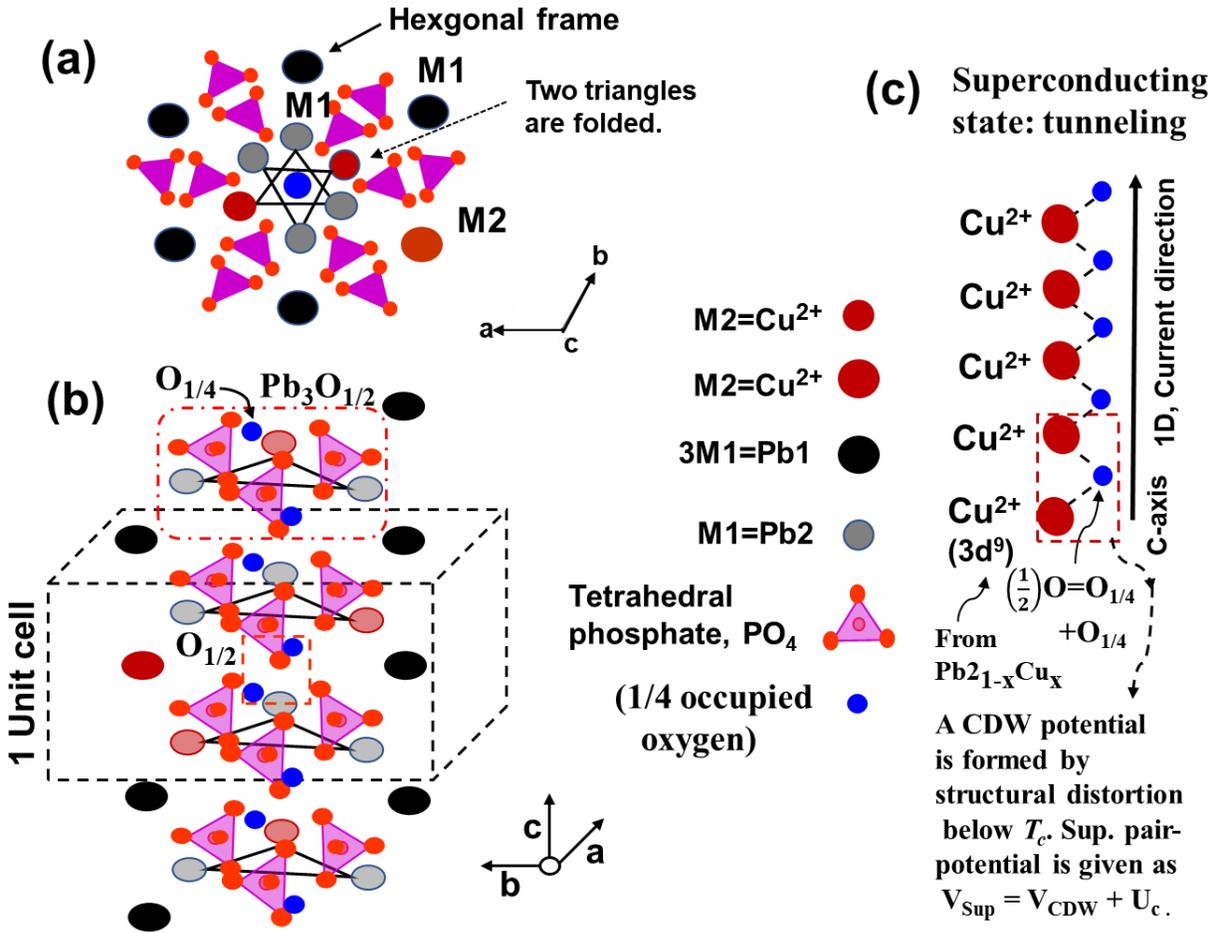

**FIG. 3.** (a) Top view of Cu-doped LA, $Pb_{10-x}Cu_x(PO_4)_6O$, to c-axis. Inside, hexagonal six (M1+M2)s corresponding to (Pb2+Cu)s (2 indicates the number of classifications of Pb(2) in cif.) are folded by two layers composed of $Pb_{3-y}Cu_yO_{1/2}$ expressed as a triangle, [Fig. 3(b)]. (b) Side view of the Cu-doped LA. 1 unit cell is also displayed. (c) In the superconducting state, layout to explain a one-dimensional superconducting chain along the c-axis. The CDW is the charge-density-wave and $V_{CDW}<0$ is the CDW potential. $V_{Sup}<0$ is a potential containing the superconducting carrier.[7] $U_c > 0$ is the critical on-site repulsive Coulomb energy in the BR picture.[17] This was re-drawn with Fig. 4 in Ref. 5.

## C. LK-99® : Meissner effect

Figures 4(a) and 4(b) illustrate the temperature-dependent diamagnetic susceptibilities of zero-field cool (ZFC) and field cool (FC) in sample 2 (obtained in a quartz vessel with low doping of lead apatite) and sample 3 (manufactured using higher-purity raw materials). The measurements were conducted in a temperature range of -73.15℃(200K) to 126.85℃(400K) and involved the observation of the Meissner effect, expelling the magnetic field. Moreover, we compare the susceptibilities in Figs. 4(a) and (b) with that measured at a graphite exhibiting the diamagnetic behavior[18]. The ratios of the susceptibilities at a room temperature of 20ºC are given as $R_1$=(-1.20 × $10^{-1}$ emu/g, Fig. 4a)/(-22 × $10^{-6}$ emu/g) ≈ 5450 and $R_2$=(-5.0 × $10^{-4}$ emu/g, Fig. 4b)/(-22 × $10^{-6}$ emu/g) ≈ 22.7. The high ratios cannot be explained except for the fact that the samples have a superconducting phase.

Fig. 4(c) shows the levitation phenomenon measured at room temperature and atmospheric pressure for sample 4 (heat-treated with sample 2), indicating the presence of a superconducting phase at room temperature and atmospheric pressure, although the levitation is not perfect. Video of the levitation is attached as Fig. 4(d). The susceptibilities were measured for samples 2 and 3 using MPMS-Evercool at Kaist Analysis center for Research Advancement.

https://arxiv.org/abs/2307.12037

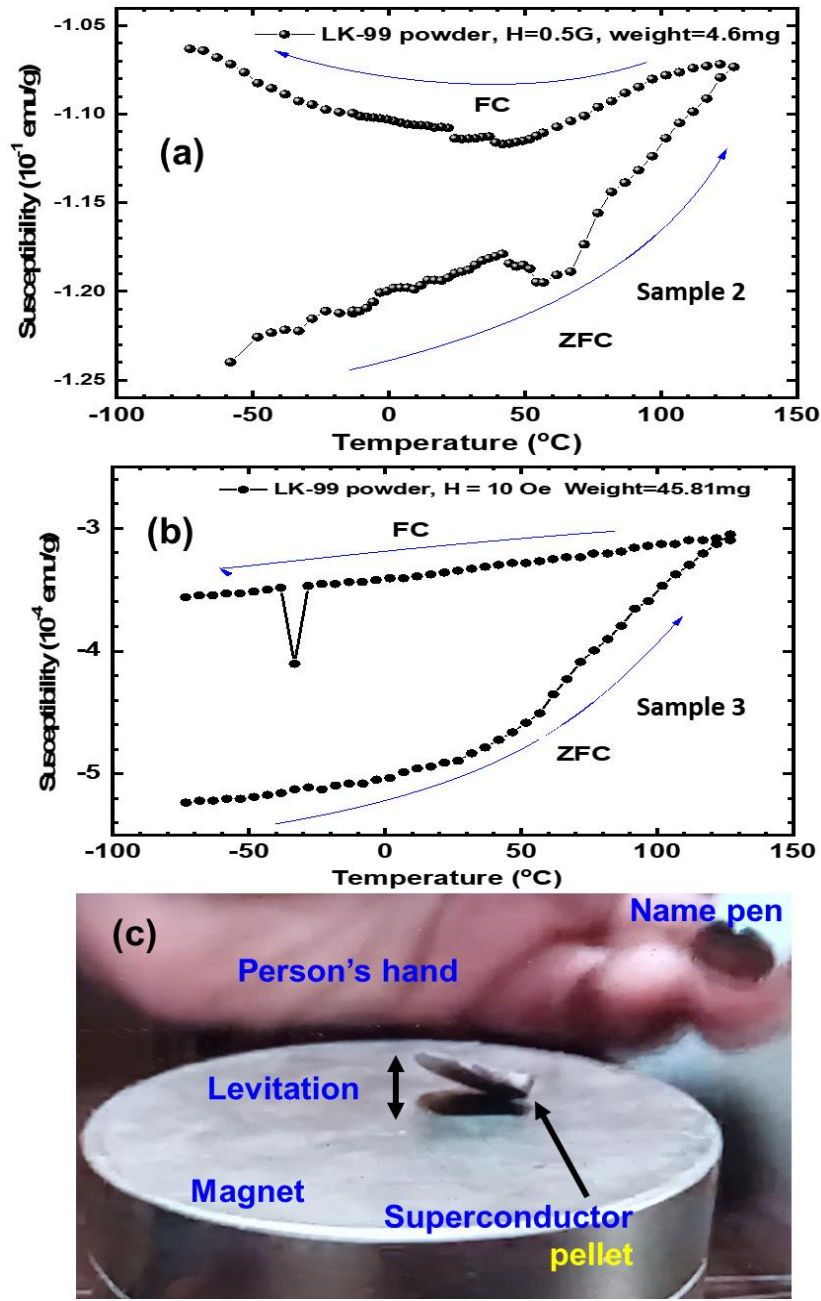

**FIG. 4.** (a) Temperature dependence of diamagnetic susceptibilities measured in samples 2 (a) and 3 (b). (c) Levitation phenomenon obtained from annealed sample 2 (sample 4). (d) A video for levitation is attached.

**D. LK-99® : Resistivity**

Figure 5 shows the temperature dependence of the resistivity of sample 2 (4.8 × 10.1 × 1.2mm) measured at 30 mA by the four-probe method. A jump in resistivity appears near $T_c$=104.8°C (377.95K). Above $T_c$, the linear characteristic of metal originating from the IMT is displayed. Below $T_c$, three different behaviors are exhibited. In the first region below red-arrow C (near 60°C), equivalent to region F in the inset of Fig. 5, the resistivity with noise signals can be regarded as zero. In the second region, indicated by the red arrows C and D (near 90°C) and corresponding to region G in the inset of Fig. 5, the resistivity of the sample monotonically increases with temperature. This indicates the occurrence of resistance, suggesting a breakdown of the superconducting energy gap as the temperature increases. At the third region between red-arrow D and $T_c$, corresponding to region H in the inset of Fig. 5, the resistivity does not transparently change with increasing temperature; however, $d\sigma/dT$ fluctuates at the final stage of the breakdown of the energy gap. In the first region, where signals resembling noise are observed, the zero-resistivity region is approximately 88% (333K/378K) of $T_c$ under Kelvin units. This is approximately three times larger than the typical value of about 30% observed in low $T_c$ superconductors. The presence of noise in the zero-resistivity region is often attributed to phonon vibrations at higher temperatures, as indicated by region F in inset 5; the signals like noise were directly measured[5]. The presence of the zero-resistivity region is evidence of an *s*-wave superconductor because a node-type superconductor with no-gap metal at node, such as $d_{x^2-y^2}$ pairing symmetry has no zero-resistivity region due to increase of metal resistivity with increasing temperature.[10,11]

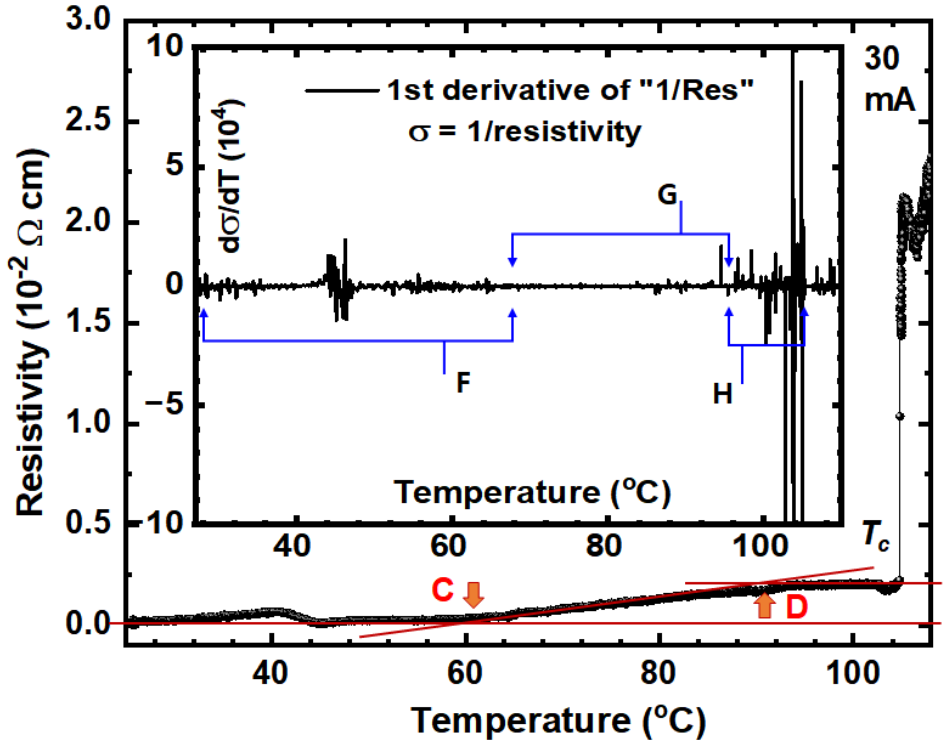

**FIG. 5.** Temperature dependence of resistivity. Inset shows $d\sigma/dT=d(1/resistivity)/dT$, regarded as the density of states (DOS). The temperature dependence of $d\sigma/dT$ is interpreted as that of DOS. Resistivity figure was re-drawn with Fig. 6(a) in Ref. 5.

**E. LK-99® : I-V characteristics**

Figure 6(a) shows the temperature dependence of an I-V curve measured in sample 1. Above $T_c$, the linear characteristic of a metal is displayed. As the temperature increases, $T_{c,\ current}$ decreases. Moreover, the magnitude of the jump of $T_{c,\ current}$ decreases with increasing current. In particular, at 105°C, the jump is very small. This indicates that the magnitude of the jump does not monotonically decrease. This is known as exponential decrease in a gap-nogap transition material.[19] We conducted a detailed analysis of the current-voltage (I-V) curve below $T_{c,\ current}$, measured at 25°C, focusing on the logarithmic y-axis. The curve can be divided into several regions: I, J, K, L, M, and N. In regions I, J, and K, we observe an increase in resistance, indicating the breaking of the superconducting energy gap due to Joule heating when the current exceeds a certain threshold. The curve is obvious, as indicated by red-arrow O in an I-V curve measured at 45°C. More high current of region L accelerates the breaking, which can be interpreted as an avalanche region. At region M of much higher current, voltage does not almost change. This indicates that the density of states (DOS)=$dI/dV$ is constant near $T_{c,\ current}$, suggesting that the pairing symmetry is *s*-wave, [Figs. 6(b) and 6(c)]. If the superconducting gap has a node, $dI/dV$ near $T_{c,\ current}$ should increase.[10] After the jump, region N is a metal with the linear Ohmic characteristic. The transition between the superconducting gap and the nogap metal is the gap-nogap transition [Fig. 6(a)]. The transition is almost identical to the characteristic of the IMT.[20,21] Moreover, the fluctuation of regions I, J, and K, [Fig. 6(c)], is owing to temperature inhomogeneity caused by Joule heat generated by both current and heat applied by the temperature controller. Fig. 6(d) shows the magnetic field dependence of sample 1. Above the jump of $T_c$, the I-V curve exhibits Ohmic characteristics, indicating a metallic behavior. As the magnetic field increases, $T_{c,\ current}$, decreases, exhibiting the typical characteristic of a superconductor. Fig. 6(e) exhibits a maximum $T_c$ of 127°C in heat-treated sample 1. Because the jump remains, we deduce that $T_c$ exceeds 127°C. Above $T_c$, the linear characteristic of a metal is displayed. A summary of the results is shown in the phase diagram [Fig. 7].

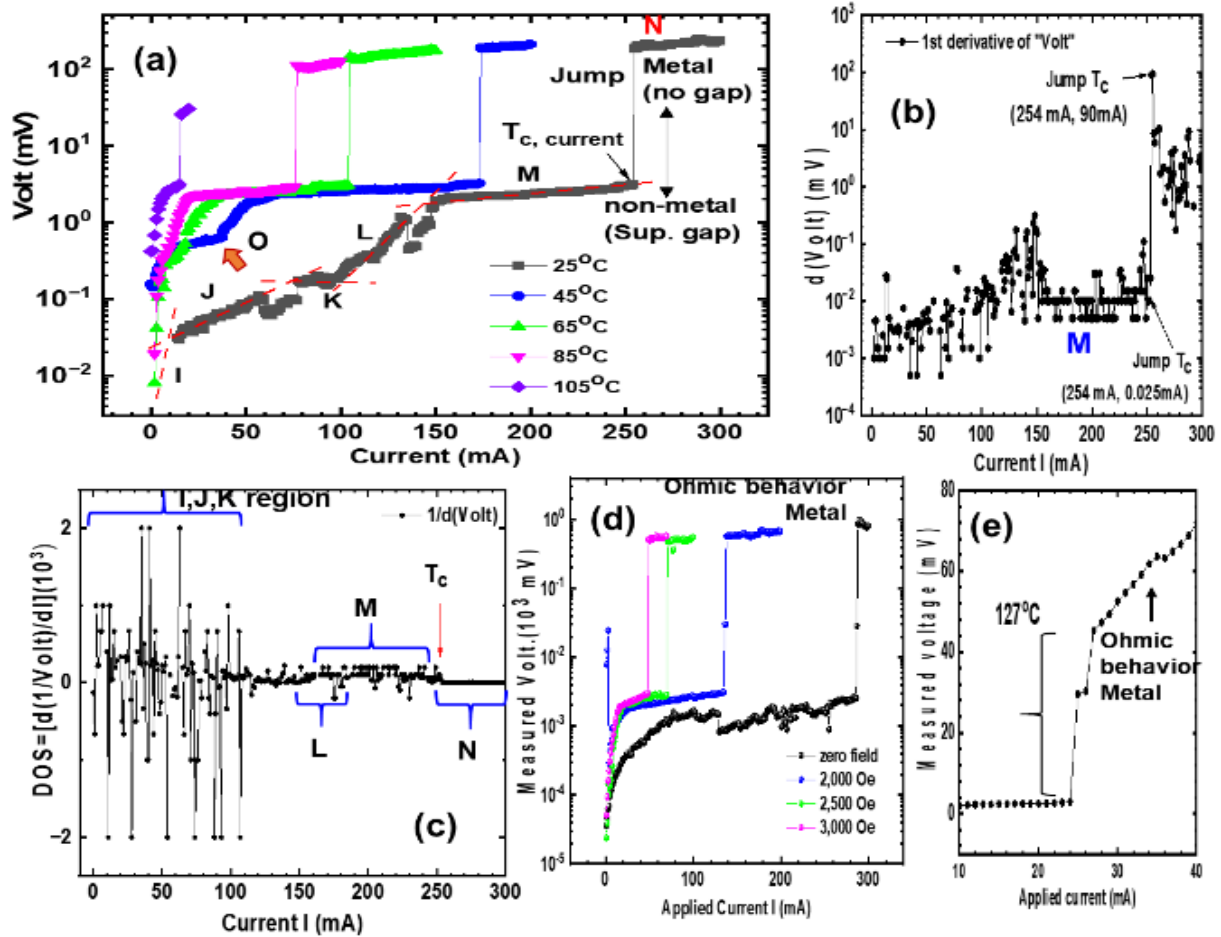

**FIG. 6.** (a) The temperature dependence of an I-V y-axis log curve, obtained through a method measuring voltage with applying current. (b) Differential curve of data at 25°C. (c) DOS = $d\sigma/dI$, the derivative curve of conductance. $dI$=1mA is constant. Regions I, J, K and M present the characteristic of *s*-wave symmetry. (d) Magnetic field dependence of I-V curves. (e) 127°C is a temperature below a $T_c$ owing to temperature limitation of the measurement system.

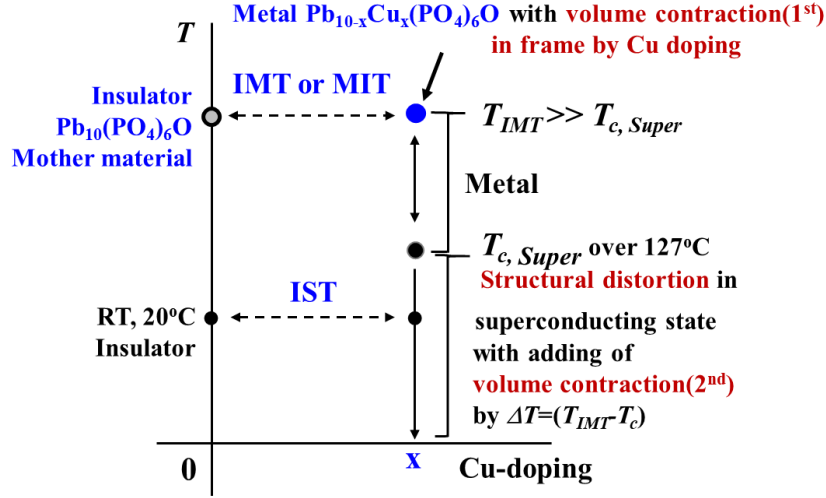

**FIG. 7.** Phase diagram of $Cu^{2+}$-doped LA $Pb_{10-x}Cu_x(PO_4)_6O$. IMT is the insulator-to-metal transition and MIT is the metal-to-insulator transition. IMT and MIT follow the same concepts, indicating a gap-nogap transition. IST is an insulator-superconductor transition (gap-gap transition), indicating a change of electric characteristics in the same gap structure.

## III. POSSIBLE MECHANISM OF SUPERCONDUCTIVITY

### A. LK-99® : One-dimensional metal

The mother material of the LA structure, as shown in Fig. 3, is disassembled by the following chemical formula,

$Pb_{10}(PO_4)_6O \equiv [Pb1_4]^F[Pb2_6]^T(PO_4)_6O$

$= [Pb1_4(PO_4)_{8/3}]^F + [Pb2_6(PO_4)_{10/3}O]^T$,

$= [Pb1_4(PO_4)_{8/3}]^F + [Pb2_3(PO_4)_{5/3}O_{1/2} + Pb2_3(PO_4)_{5/3}O_{1/2}]^T$,

$= [Pb1_4(PO_4)_{(2+2/3)}]^F + [Pb2_3(PO_4)_{(1+2/3)}O_{1/2} + Pb2_3(PO_4)_{(1+2/3)}O_{1/2}]^T$, ---- (1)

where F names the frame part, T denotes a part of the tunnel along the c-axis, and $O_{1/2}$ is composed of $O_{1/4} + O_{1/4}$. This highly stable structure consists of two layers of frames with an interior region.[22,23,24] Formula (1) is characterized by a one-dimensional bonding of Pb2 and $O_{1/2}$ in the T part. When parts of the Pb1 (meaning the Pb(1) site in LA cif.) and the Pb2 (meaning the Pb(2) position in LA cif.) sites are randomly substituted by $Cu^{2+}$ ($3d^9$) (one hole) elements, chemical formula of the $Cu^{2+}$-doped LA structure is expressed as follows:

$Pb_{10-x}Cu_x(PO_4)_6O = [Pb_{4-a}Cu_a(PO_4)_{8/3}]^F + [Pb_{6-b}Cu_b(PO_4)_{10/3}O]^T$, where $a+b=x$

$= [Pb_{4-a}Cu_a(PO_4)_{8/3}]^F + [Pb_{3-c}Cu_c(PO_4)_{5/3}O_{1/2} + Pb_{3-c}Cu_c(PO_4)_{5/3}O_{1/2}]^T$, where $c=b/2$, a and b $\neq 0$

$= [Pb_{4-a}Cu_a(PO_4)_{(2+2/3)}]^F + [Pb_{3-c}Cu_c(PO_4)_{(1+2/3)}O_{1/2} + Pb_{3-c}Cu_c(PO_4)_{(1+2/3)}O_{1/2}]^T$. ---- (2)

Hole doping quantities are determined as the outmost orbitals of $-3a$ ($=2(4-a)+(-1)a+(-3)(2+2/3)$) in the frame part and $-3c$ ($=2(3-c)+(-1)c+(-3)(1+2/3)+(-2)(1/2)$) in the T part. When IMT occurs by $Cu^{2+}$ doping, IMT is generated at both the F and the T parts. Generally, because an insulator has a larger volume than that of a metal, the IMT is accompanied by a volume contraction without the structural phase transition. The substitutions at the Pb1 sites, as indicated by the reddish-brown color in Fig. 3(a), induce a volume contraction($1^{st}$) in the frame through the IMT($1^{st}$).[16] Moreover, substitutions at the Pb2 sites lead to an IMT($2^{nd}$) in the T part. The evidence of the volume contraction is shown in Fig. 2(b). If the doping quantities of $a$ and $c$ in formula (2) differ, the extent of the IMT and the magnitude of volume contraction may vary. If $a>c$, below a critical doping level, the volume contraction($1^{st}$) in the frame may be larger than the volume contraction($2^{nd}$) in the T part.

The outmost orbital of Pb is $6s^2 6p^2$. One electron in the $6p^2$ orbital of Pb2 is coupled to the O of PbO, whereas the other contributes to the bonding of tetrahedral phosphate, $PO_4$, [formulas (1) and (2)]. Consequently, the inside structure is stabilized. Lone-pair electrons of $6s^2$ in Pb play an important role on the phase transition; the $6s^2$ electrons are spherically symmetric and displaced structure relative to the position of Pb atoms and contributing to polar interaction,.[24] Pbs and Cu are bonded through tetraphosphates. The metal phase comes from $Cu^{2+}(3d^9)$ with one hole substituted, as shown in Fig. 3. A metal line such as $Cu^{2+}(3d^9)-O_{1/2}-Cu^{2+}(3d^9)$ along the c-axis is formed. This is referred to as the hole-driven insulator-to-metal transition (IMT).[20,21,25] Hence, $Pb2_{3-y}Cu_y(O_{1/2} = O_{1/4} + O_{1/4})$ structures are generated and the oxygen is located at slightly higher or slightly lower position than the $Pb2_{3-y}Cu_y$ structure, [blue balls in red dot box in Fig. 3(b)]. The nearest two oxygens (blue balls) at the $O_{1/4}$ position in $O_{1/2}$ of 1-D channel, as shown in the red-dot box of unit cell of Fig. 3b, are generally repulsive. In metal case, when one oxygen of the $O_{1/4}$ vibrates (the distance between $Pb2_{3-y}Cu_y$ and $O_{1/4}$ expands and contracts), the other anti-vibrates (the distance contracts and expands, respectively). This indicates that oxygen breathes; the average distance between $Pb2_{3-y}Cu_y$ and $O_{1/4}$ is same. One unit cell has two $Pb2_{3-y}Cu_yO_{1/2}$ structures [Fig. 3(b)]. In the metal state, the hole carrier of $Cu^{2+}$ in $Pb2_{3-y}Cu_y(3d^9)$ flows through the conduction band formed by $Cu^{2+}-Cu^{2+}$, and oxygens in the $Cu^{2+}$ (in lower $Pb2_{3-y}Cu_yO_{1/2}$) –> $O_{1/2}$ –> $Pb2_{3-y}Cu_y$ (in higher $Pb2_{3-y}Cu_yO_{1/2}$) structure breath along c-axis, [Fig. 3(c)]. In the superconducting state, carriers of $Cu^{2++}(3d^9)$s at nearest neighbor sites form a bi-polaron in a superconducting bound state ($V_{Super\ potential} = V_{CDW\ potential} + U_{c,\ critical\ on\text{-}site\ Coulomb\ energy}$), where charge-density-wave state (CDW) has the structure of long and short distances in ($Pb2_{3-y}Cu_y\text{-}O_{1/4}$ or $Pb2_{3-y}Cu_y\text{-}O_{1/2}$) between nearest neighbor Cu sites formed by breathing mode distortion (stopping breath) between oxygens in the $Pb2_{3-y}Cu_yO_{1/2}$ structures). Then, the distance between $Pb2_{3-y}Cu_y$ and $O_{1/4}$ differs, which is the structural distortion and the on-site Coulomb interaction, $U$, between carriers in metal is changed to $U_c$. The formation of the superconducting bound state is facilitated by the enhanced on-site Coulomb repulsive interaction, resulting from both the structural contraction (1st) at the frame of the LA induced by $Cu^{2+}$ doping and the structural distortion accompanied by volume contraction (2nd) owing to the temperature difference ($\Delta T = T_{IMT} - T_c$) at $T_c$. These factors contribute to superconducting condensation, as depicted in the phase diagram shown in Fig. 7.[7] The bi-polaron is able to tunnel through a barrier between two $Pb2_{3-y}Cu_yO_{1/2}$ structures in a one-dimensional chain along the c-axis, where the Pb2 in the lower $Pb2_{3-y}Cu_yO_{1/2}$ is connected to $O_{1/2}$ and to Cu in the higher $Pb2_{3-y}Cu_yO_{1/2}$ (as shown in Figure 3(c)). Additionally, the first term of the frame part in formula (2) can exhibit superconductivity owing to the IMT; however, because it is not one-dimensional, its Tc will be much smaller compared to that of the T part in Formula (2). The underlying reasons for this behavior will be further explained in the subsequent section.

**B. LK-99® : Strong correlation**

As for a room-temperature-$T_c$ mechanism, the MLA structure inducing the IMT is characterized by volume contractions(1st and 2nd) by substituting Pb with Cu. A known theory to explain room-temperature superconductivity, including structural volume contraction, is BR-BCS, suggesting that divergence of the DOS, caused by on-site Coulomb repulsive interaction $U$ increased by volume contraction, increases superconducting $T_c$ over room temperature in the BR-BCS $T_c$.[7,26] When dimensionality is considered, 1-dimension(D) DOS(=$N(0)$) is proportional to $(m^*/E^*)^{0.5}$ with the divergence to $\rho$, where an effective mass of carrier, $m^* \equiv m/[1-(U/U_c)^2] = m/[1-\rho^4\kappa^2]$, and kinetic energy, $E^* \equiv E_k(1-U/U_c)^2 = E_k(1-\rho^2\kappa)^2$, $U/U_c = \rho^2\kappa$ with percolation to increasing $\rho$ as a function of doping $x$ and correlation strength $\kappa \approx 1(\neq 1)$ generating the maximum number of excited carriers,[17] band-filling factor $0 < \rho \leq 1$ are defined.[7] The effective 2D-DOS$^*$=2D-DOS$_{(non-interaction\_BCS)}/[1-\rho^4\kappa^2]$ with the divergence to $\rho$ is given because of 2D-DOS$^* \propto \underline{m}^*$. 3D-DOS$^* \propto (m^*)^{1.5}(E^*)^{0.5}$ has the divergence to $\rho$. 1D-DOS$^*$=1D-DOS$_{(non-interaction\_BCS)}/[(1-\rho^4\kappa^2)^{0.5}(1-\rho^2\kappa)]$ where $\kappa \approx 1(\neq 1)$. 1D-DOS$^*$ is larger than 2D-DOS$^*$ and 3D-DOS$^*$ in the same condition [Fig. 8(a)]. When 1D-DOS$^*$=$N^*(0)$ is applied to electron-phonon coupling $\lambda^* \equiv N^*(0)V = A\lambda_{non-interaction}$ in the BCS-$T_c$ formula, where $A=1/[(1-\rho^4\kappa^2)^{0.5}(1-\rho^2\kappa)]$ and $\kappa \approx 1(\neq 1)$ is given, $V$ is the attractive electron-phonon potential and is known as about 0.2 eV[28], $T_c$ will increase above room temperature when $\rho$ approaches one, on the basis of the BR-BCS $T_c$[7], [Figs. 8(b) and 8(c)]. In calculations of $T_c$, the Debye temperature is used as $T_{Debye}$=840 K[27] determined in CuO regarded as 1D structure [Fig. 3(c)]. Figs. 8(b) and 8(c) show that 1D $T_c$s are higher than those at 2D and 3D although values of the band-filling factor are smaller, which indicates the coupling is stronger. The BR-BCS theory uses a bi-polaron strongly coupled by the attractive charge-density-wave potential (short-range electron-phonon interaction), instead of the Cooper pair defined as coupling of excited two electrons bound by the attractive screened long-range electron-phonon (atom) interaction.[7,28] Moreover, the general mechanism of superconductivity in the Fermi system with strong repulsive interaction explained in the 2D system that superconducting $T_c$ reaches about 10% of Fermi temperature (if $T_{Fermi} \approx 10{,}000$ K is assumed),[8] although this mechanism is not involved in volume contraction. The theory of hole superconductivity suggested much higher $T_c$ than that proposed in BCS theory.[25] In addition, one-dimensional superconductivity and $T_c$ enhancement were already disclosed.[29,30,31,32]

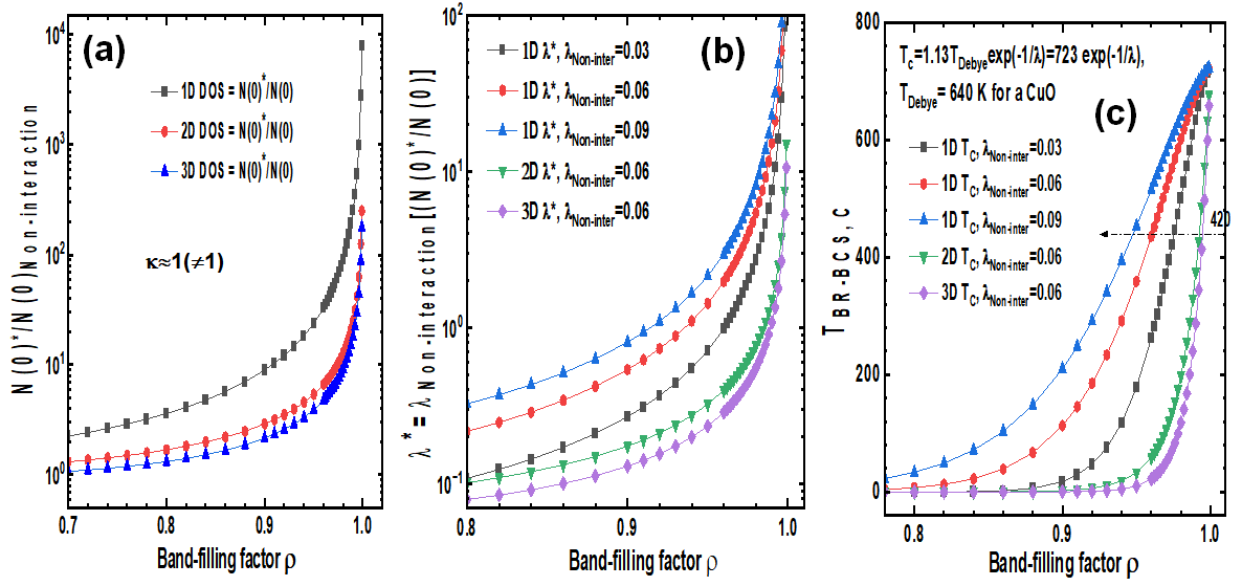

**FIG. 8.** (a) A ratio of an effective density of states (DOS)=N(0)* and N(0), in cases of the presence (or absence) of on-site Coulomb repulsive interaction, respectively is shown. The ratio is expressed as a function of band-filling factor of $\rho$. One dimensional(D) DOS is much larger than those of 2D and 3D. The formula was given in the main text. (b) $\lambda^* = N(0)^*V$ is an effective electron-phonon coupling constant. $V \approx 0.2$ eV is the electron-phonon coupling potential given in the Cooper paper[28]. $\lambda_{Non-interaction} = N(0)V$ is constant when the Coulomb interaction does not exist. Generally, a value of $\lambda_{Non-interaction}$ is very small.[7] Because it is not determined, several values are selected for comparing the values. (c) The BR-BCS $T_c$s are displayed as a dimensional function in cases of several $\lambda_{Non-interaction}$s. One-D $T_c$ is much bigger than those of 2 D and 3D. Here, a Debye temperature of 640K calculated in CuO is used,[27] because the 1D metal line is formed by $Cu^{2+}$ –> $O_{1/2}$ –> $Cu^{2+}$.

## IV. CONCLUSION

We have successfully developed a method for synthesizing a superconducting material with an MLA structure, exhibiting both one-dimensional characteristics and a $T_c$ above room temperature at atmospheric pressure. The presence of superconductivity was confirmed through observations of the levitation phenomenon and analysis of zero resistivity. The unique features of the MLA structure include volume contraction (1st) resulting from substituting Pb with Cu. On-site Coulomb repulsive interaction increased by volume contractions(1st and 2nd) may cause the superconducting phenomenon, as mentioned in the BR-BCS theory explaining the room-temperature $T_c$. Furthermore, the room-temperature superconductor opens up possibilities for high-performance superconducting wires and magnets operating at room temperature, with potential applications in energy transmission, transportation, and scientific research.

https://arxiv.org/abs/2307.12037

ACKNOWLEDGEMENTS

We acknowledge late Prof. Chair Tong-seek for initiating research of a 1-dimensional superconductor of over room temperature at atmospheric pressure. In particular, his enthusiasm on superconductor study impressed many researchers. Moreover, we thank Mr. Ki Se-woong, Mr. Lee Byungkyu (CEO of ProCell Therapeutics, Inc.), Mr. Yoon Sang-ok (Chairman of FINE Inc.) so much for financial supporting, and Bang Jaekyu and Kim Gyeonkcheol so much for wholeheartedly sharing the burdens and difficulties in this investigation. This research was primarily supported by research-and-development funds from Quantum Energy Research Centre Inc.. SQUID measurements were supported by the National Research Foundation of Korea grant funded by the Korea government(MSIT) (No. 2019R1I1A01059675) and Korea University Grant (Projects of an author, Young-Wan, Kwon taking charge of SQUID measurements). We thank Prof. Mumtaz Qazilbash for valuable comments. An author, Hyun-Tak Kim (H. T. Kim),'s knowledge on mechanisms of both superconductivity and the metal-insulator (gap-nogap) transition highly contributed to writing the mechanism part. The knowledge was acquired over 20 years by processes of performing national projects including project [Grant 2017-0-00830] funded by Institute for Information and Communications Technology Promotion (IITP) in MSIT of Korea government in ETRI. H. T. Kim left ETRI on Nov. of 2022.

AUTHOR DECLARATIONS

**Conflict of Interest**

The authors have no conflicts to disclose.

**Author Contributions**

Sukbae, Lee: Conceptualization(lead); Data curation(equal); Funding acquisition(lead); Investigation(equal); Methodology(equal); Project administration(lead); Resources(equal); Software(equal); Supervision(lead); Validation(equal); Visualization(support); Writing – original draft(equal); Writing – review & editing(equal), Ji-hoon, Kim: Conceptualization(equal); Data curation(equal); Formal analysis(equal); Investigation(equal); Methodology(equal); Project administration(equal); Software(equal); Supervision(equal); Validation(equal); Visualization(equal), Sungyeon, Im: Data curation(support); Funding acquisition(equal); Resources(equal); Supervision(equal); Validation(equal) SooMin, An: Data curation(support); Funding acquisition(support); Investigation(support); Project administration(support); Resources(support); Validation(support); Writing – original draft(support); Keun Ho, Auh: Funding acquisition(support); Methodology(equal); Project administration(support); Supervision(equal); Writing – original draft(lead). Hyun-Tak Kim analyzed $s$-wave symmetry, and made room-tem.-$T_c$ mechanism including metal-insulator transition and CDW structural distortion through structure analysis and brought the levitation into this paper, wrote this manuscript with authors.

DATA AVAILABILITY

The data that support this study are available from the corresponding authors upon reasonable request.

https://arxiv.org/abs/2307.12037

https://arxiv.org/abs/2307.12037